# INFLATIONARY RSII MODEL WITH A MATTER IN THE BULK AND EXPONENTIAL POTENTIAL OF TACHYON FIELD[†]

*UDC 524.82, 539.120.52*

**Marko Stojanović[1], Neven Bilić[2], Dragoljub D. Dimitrijević[3], Goran S. Đorđević[3], Milan Milošević[3]**

[1]Faculty of Medicine, University of Niš,
Bulevar Dr Zorana Đinđića 81, 18000 Niš, Serbia
[2] Division of Theoretical Physics, Rudjer Bošković Institute,
Bijenička cesta 54, 10000 Zagreb, Croatia
[3] Department of Physics, Faculty of Science and Mathematics, University of Niš,
Višegradska 33, 18000 Niš, Serbia Serbia

**Abstract**. *In this paper, we study a tachyon cosmological model based on the dynamics of a 3-brane in the second Randall-Sundrum (RSII) model extended to include matter in the bulk. The presence of matter in the bulk changes the warp factor which leads to the modification of inflationary dynamics. The additional brane behaves effectively as a tachyon. We calculate numerically observation parameters of inflation: the scalar spectral index ($n_s$) and the tensor-to-scalar ratio (r) for the exponential potential of the tachyon field.*

## 1. INTRODUCTION

The theory of cosmological inflation provides a way to solve flatness, horizon and most of the other problems in standard cosmology by the rapid expansion of the universe (Guth, 1981; Linde, 1982; Starobinsky, 1979). Also, inflation generates density perturbations which are responsible for creating a large-scale structure in the universe (Mukhanov and Chibisov, 1981).

Received April 4th, 2019; accepted April 25th, 2019
**Corresponding author**: Marko Stojanović
Faculty of Medicine, University of Niš, Dr. Zorana Đinđića Blvd. 81, 18000 Niš, Serbia
E-mail: marko.stojanovic@pmf.edu.rs
[†] Acknowledgement: This work has been supported by the ICTP - SEENET-MTP project NT-03 Cosmology-Classical and Quantum Challenges. The work of N. Bilić has been supported by the H2020 CSA Twinning project No. 692194, "RBI-T-WINNING". N. Bilić, M. Milošević and G. Djordjevic acknowledge the support provided by the STSM CANTATA-COST grant. D.D. Dimitrijević, G.S. Djordjevic, M. Milošević and M. Stojanović acknowledge the support provided by the Serbian Ministry for Education, Science and Technological Development under the projects No. 176021, No. 174020 and No. 176003.

Although there is strong evidence of inflation (e.g., CMB radiation), the precise mechanism of the process is still unknown. Various models of inflation have been proposed and they can be classified according to the corresponding underlying type of Lagrangian, the type of the potential, etc. Our consideration is based on the Lagrangian of the Dirac-Born-Infeld (DBI) form, arising from the string field theory (Dimitrijevic et al., 2016; Djordjevic et al., 2016; Sen, 1999)

$$\mathcal{L} = -V(\Theta)\sqrt{1 - g^{\mu\nu}\Theta_{,\mu}\Theta_{,\nu}}, \qquad (1)$$

where $\Theta$ denotes the tachyon scalar field and $V(\Theta)$ its potential, which satisfies specific conditions (Steer and Vernizzi 2004)

$$V(0) < \infty, \quad dV/d\Theta(\Theta > 0) < 0, \quad V(|\Theta| \to \infty) \to 0. \qquad (2)$$

The Randall-Sundrum (RSI and RSII) models belong to the class of brane-world models, which are based on the idea that ordinary matter (standard model particles) is confined on a hypersurface, called brane, located in the higher dimensional spacetime, called bulk (Randall and Sundrum, 1999a; Randall and Sundrum, 1999b). Although the RSI model was proposed as a solution to the hierarchy problem (unnatural discrepancy between the strength of gravity and those of the other forces) in particle physics, its extension by introducing one extra dimension compactified on a circle, the model has shown a significant potential for strong implication in high-energy astrophysics, black holes and cosmology. The original RSI model consists of two D3-branes, with opposite tensions, in the five-dimensional bulk which is a slice of the anti de Sitter space (AdS$_5$) (Randall and Sundrum, 1999a). Here, only gravity interaction can freely propagate in the bulk (Antoniadis et al., 1998; Arkani-Hamed et al., 1998).

The second Randall-Sundrum (RSII) model effectively consists of only one brane with positive tension (because the negative tension brane is pushed off to infinity) and provides a mechanism for localizing gravity without compactification of the extra dimension (Randall and Sundrum 1999b). One additional D3-brane moving in the AdS$_5$ bulk behaves effectively as a tachyon with the potential $V(\Theta) \propto \Theta^{-4}$ (Bilić and Tupper, 2014). The model can be extended to include matter in the bulk (Bilić et al., 2017a; Bilić et al., 2017b). In this paper, we study tachyonic inflation driven by the dynamics of a D3-brane, from an extended RSII model, and calculate numerically observation parameters of inflation.

The remainder of the paper is organized as follows. In Section 2 we introduce the tachyon as a dynamical brane. In Section 3 we derive the field equations in a covariant Hamiltonian formalism. The slow-roll regime is discussed in Section 4. Section 5 is devoted to the observational cosmological parameters. The numerical method and results obtained for exponential potential were presented in Section 6. Section 7 concludes this study.

## 2. DYNAMICAL BRANE AS TACHYON

We consider a (3+1)-dimensional dynamical D-brane in the (4+1) dimensional bulk spacetime with coordinates $X^a$, $a = 0,1,2,3,4$. The points on the brane are parametrized by $X^a(x^\mu)$, $\mu = 0,1,2,3$, where $x^\mu$ are the (local) coordinates on the brane. The line element of AdS$_5$ space is

$$ds^2_{(5)} = G_{ab}dX^a dX^b = e^{-2|y|/\ell}\eta_{\mu\nu}dx^\mu dx^\nu - dy^2, \tag{3}$$

where $\ell$ is the AdS curvature radius and $x^5 = y$ is the extra dimension (Bilić and Tupper, 2014). It is suitable to work in conformal coordinates with the line element

$$ds^2_{(5)} = \frac{1}{\chi^2(y)}(g_{\mu\nu}dx^\mu dx^\nu - dy^2). \tag{4}$$

The action of the brane is given by Johnson (2003)

$$S_{br} = -\sigma \int d^4x\, e^{-\phi_d}\sqrt{-\det(\hat{g}^{(ind)} + \mathcal{B})}, \tag{5}$$

where $\sigma$ is the brane tension, $\phi_d$ is the dilaton field, $\hat{g}^{(ind)}$ is the tensor of the induced metric on the brane and the object $\mathcal{B}$ denotes an antisymmetric tensor field. Using Gaussian normal parametrization, the induced metric can be expressed in the form

$$g^{(ind)}_{\mu\nu} = \frac{1}{\chi^2(\Theta)}(g_{\mu\nu} - \Theta_{,\mu}\Theta_{,\nu}). \tag{6}$$

Neglecting the dilaton scalar field and the antisymmetric tensor field, the action for the dynamical brane takes the form of the action for the tachyon scalar field of the DBI-type (Bilić et al., 2017a; Bilić et al., 2017b; Dimitrijević et al., 2018)

$$S_{br} = S_{tach} = -\sigma \int d^4x\, \frac{1}{\chi^4(\Theta)}\sqrt{1 - g^{\mu\nu}\Theta_{,\mu}\Theta_{,\nu}}, \tag{7}$$

with the Lagrangian density of the form (1), and the tachyon potential

$$V(\Theta) = \frac{\sigma}{\chi^4(\Theta)}. \tag{8}$$

A more general tachyon potential could be obtained from a more general bulk geometry.

### 3. FIELD EQUATIONS

The treatment of our system is performed in the relativistic Hamiltonian formalism. Under the assumption that the spacetime on the observer's brane is a flat (FRW) universe, the Hamiltonian density associated with (7) is

$$\mathcal{H} = \frac{\sigma}{\chi^4(\Theta)}\sqrt{1 + \frac{1}{\sigma^2}\Pi_\Theta^2 \chi^8}, \tag{9}$$

where $\Pi_\Theta$ is the magnitude of the conjugate momentum associated with the tachyon scalar field $\Theta$, defined as $\Pi_\Theta^\mu = \partial \mathcal{L}/\partial \Theta_{,\mu}$. From the Hamilton equations, we obtain

$$\dot{\Theta} = \frac{\partial \mathcal{H}}{\partial \Pi_\Theta}, \tag{10}$$

$$\dot{\Pi}_\Theta + 3H\Pi_\Theta = -\frac{\partial \mathcal{H}}{\partial \Theta}, \tag{11}$$

where the overdot represents a derivative with respect to the coordinate time $t$. The Friedman equations are of the form (Bilić et al., 2017b; Dimitrijević et al., 2018)

$$H = \frac{\dot{a}}{a} = \sqrt{\frac{8\pi G_N}{3}\mathcal{H}\left(\frac{\chi_{,\theta}}{k} + \frac{2\pi G_N}{3k^2}\mathcal{H}\right)}, \tag{12}$$

$$\dot{H} = -4\pi G_N(\mathcal{H}+\mathcal{L})\left(\frac{1}{k}\chi_{,\theta} + \frac{4\pi G_N}{3k^2}\mathcal{H}\right) + \sqrt{\frac{2\pi G_N \mathcal{H}}{3k\chi_{,\theta}+2\pi G_N \mathcal{H}}}\frac{d\chi_{,\theta}}{dt}, \tag{13}$$

where $k$ is a ratio between the four-dimensional Newton constant $G_N$ and the five-dimensional constant $G_5$. Due to the existence of an extra dimension, Eqs. (12) and (13) are obviously different from the Friedman equations obtained in standard cosmology.

In order to express all relevant functions in terms of dimensionless quantities, it is convenient to rescale the time $t = \tau/k$ and introduce the dimensionless Hubble expansion rate $h$, tachyon field $\theta$ and its conjugate momentum $\pi_\theta$

$$h = \frac{1}{k}H, \quad \theta = k\Theta, \quad \pi_\theta = \frac{1}{\sigma}\Pi_\theta. \tag{14}$$

For the scalar field minimally coupled with gravity, the Lagrangian and the Hamiltonian densities are identified with the pressure and energy density, respectively. The dimensionless pressure and energy density are

$$p = \frac{1}{\sigma}\mathcal{L}, \quad \rho = \frac{1}{\sigma}\mathcal{H}. \tag{15}$$

It is also convenient to introduce the dimensionless coupling constant $\kappa$ (Bilić et al., 2017b)

$$\kappa^2 = \frac{8\pi G_N}{k^2}\sigma. \tag{16}$$

Now, the system of dimensionless equations corresponding to Eqs. (10)-(12) becomes

$$\dot{\theta} = \frac{\pi_\theta}{\rho}, \tag{17}$$

$$\dot{\pi}_\theta = -3h\pi_\theta + \frac{4}{\chi^5\sqrt{1+\chi^8\pi_\theta^2}}\chi_{,\theta}, \tag{18}$$

$$h = \sqrt{\frac{\kappa^2}{3}\rho\left(\chi_{,\theta}+\frac{\kappa^2}{12}\rho\right)}. \tag{19}$$

As of now, the overdot denotes a derivative with respect to the rescaled time $\tau$.

## 4. SLOW-ROLL REGIME

We use the following definition of the slow-roll parameters (Schwarz et al., 2001)

$$\epsilon_j = \frac{d \ln |\epsilon_{j-1}|}{H dt}, \qquad j \geq 1, \tag{20}$$

$$\epsilon_0 \equiv \frac{H_*}{H}, \tag{21}$$

where $H_*$ is the Hubble expansion rate at the arbitrarily chosen time. The first two parameters are

$$\epsilon_1 = -\frac{\dot{h}}{h^2}, \tag{22}$$

$$\epsilon_2 = 2\epsilon_1 + \frac{\ddot{h}}{h\dot{h}}. \tag{23}$$

The slow-roll conditions are satisfied when $\epsilon_i \ll 1$. In this regime, during tachyon inflation, the following approximations hold (Steer and Vernizzi, 2004)

$$\dot{\theta} \ll 1, \quad \dot{\pi}_\theta \ll 3h\pi_\theta. \tag{24}$$

Using the functional dependence of the Hubble expansion rate $h$, the first two parameters in the slow-roll regime are (Bilić et al., 2018b; Dimitrijević et al., 2018)

$$\epsilon_1 \simeq \frac{8\chi^2 \chi_{,\theta}^2}{\kappa^2} \left( \chi_{,\theta} + \frac{\kappa^2}{4\chi^4} - \frac{\chi \chi_{,\theta\theta}}{4\chi_{,\theta}} \right) \left( \chi_{,\theta} + \frac{\kappa^2}{12\chi^4} \right)^{-2}, \tag{25}$$

$$\epsilon_2 \simeq \frac{8\chi^2 \chi_{,\theta}^2}{\kappa^2} \left( \chi_{,\theta} + \frac{\kappa^2}{12\chi^4} \right)^{-2} \left( \chi_{,\theta} + \frac{\kappa^2}{12\chi^4} - \frac{\kappa^2 \chi_{,\theta}}{6\chi^4} \left( \chi_{,\theta} + \frac{\kappa^2}{6\chi^4} \right)^{-1} \right). \tag{26}$$

In Eq. (26) the term proportional to the second derivative $\chi_{,\theta\theta}$ was neglected. The number of e-folds is defined as

$$N = \int_{\tau_i}^{\tau_f} d\tau h, \tag{27}$$

where $\tau_i$ and $\tau_f$ are the beginning and the end of inflation, respectively. In the slow-roll regime Eq. (27) takes the form

$$N \simeq \frac{\kappa^2}{4} \int_{\chi_i}^{\chi_f} \frac{d\chi}{\chi^3 \chi_{,\theta}^2} \left( \chi_{,\theta} + \frac{\kappa^2}{12\chi^4} \right). \tag{28}$$

In order to determine the number of e-folds, the values of the function $\chi$ at the beginning and at the end of inflation must be estimated. Because the beginning of inflation is characterized with (Bilić et al., 2017b)

$$\chi_{,\theta}(\theta_i) \ll \frac{\kappa^2}{12\chi_i^4}, \quad (29)$$

the value of the first slow-roll parameter (at the beginning of inflation) can be approximated by

$$\varepsilon_1(\theta_i) \simeq 192 \frac{\chi^6(\theta_i)\chi_{,\theta}^2(\theta_i)}{\kappa^4}. \quad (30)$$

The period close to the end of inflation is characterized by $\chi_f \gg 1$, so it can be taken

$$\frac{\kappa^2}{\chi_f^4} \ll \chi_{,\theta}(\theta_i). \quad (31)$$

The first slow-roll parameter obtained at the end of inflation is expressed as

$$\epsilon_1(\theta_f) \simeq 8 \frac{\chi^2(\theta_f)\chi_{,\theta}(\theta_f)}{\kappa^2} \left(1 - \frac{\chi(\theta_f)\chi_{,\theta\theta}(\theta_f)}{4\chi_{,\theta}^2(\theta_f)}\right) \simeq 1. \quad (32)$$

We will investigate the cosmological model with the exponential potential, in which case the function $\chi$ takes the form

$$\chi(\theta) = \exp(\theta/4). \quad (33)$$

In this case the integral in Eq. (28) has an analytical solution

$$N \simeq \frac{1}{2\varepsilon_1(\theta_i)} - \frac{2}{3}. \quad (34)$$

## 5. OBSERVATIONAL PARAMETERS OF INFLATION

In order to test the model, i.e. to compare its results with the observational data, we need to calculate the values of the observable parameters: tensor-to-scalar ratio $r$ and the scalar spectral index $n_s$ defined by

$$r = \frac{\mathcal{P}_T}{\mathcal{P}_S}, \quad (35)$$

$$n_s - 1 = \frac{d\ln\mathcal{P}_S}{d\ln q}, \quad (36)$$

where $\mathcal{P}_S$ and $\mathcal{P}_T$ are the power spectra of scalar and tensor perturbations, respectively, evaluated at the horizon, i.e., for a wave-number satisfying $q = aH$. The parameters are constrained by CMB and LSS observations (Ade et al., 2016b). Using the slow-roll approximation, the expressions for $r$ and $n_s$ up to the second order in the slow-roll parameters are

$$r = 16\epsilon_{1i}\left[1 - \frac{1}{6}\epsilon_{1i} + C\epsilon_{2i}\right], \quad (37)$$

$$n_s = 1 - 2\epsilon_{1i} - \epsilon_{2i} - \left[ 2\epsilon_{1i}^2 + \left(2C + \frac{17}{6}\right)\epsilon_{1i}\epsilon_{2i} + C\epsilon_{2i}\epsilon_{3i} \right], \tag{38}$$

where $C = -0.72$. The parameters $r$ and $n_s$ depend on the coupling $\kappa$ through $\epsilon_1$ and $\epsilon_2$.

## 6. NUMERICAL RESULTS

In order to solve the system of Hamilton's equations (17) and (18), we apply a slightly modified numerical procedure developed in Ref. (Bilić et al., 2017c). For each pair of randomly chosen parameters $N$ and $\kappa$ we have to determine initial conditions for the tachyon field, i.e. the initial conditions for $\chi$ function. The initial conjugate momentum is taken to be zero.

The system of first-order differential equations is evolved numerically using the Runge-Kutta method, from $\tau_0 = 0$ up to some large arbitrarily value $\tau_{max}$ which can provide the end of inflation ($\epsilon_1 \simeq 1$). The evolution of the Hubble parameter is determined by Eq. (19) and the evolution of the slow-roll parameters are obtained using Eqs. (22) and (23).

It is worth noting that at the end of inflation, at the time $\tau_f$, the value of the function $\chi(\theta_f)$ is determined by Eq. (32). To determine $\chi(\theta_i)$, the integral in Eq. (27) is transformed into a differential equation

$$dN = hd\tau, \tag{39}$$

that is solved simultaneously with Hamilton's equations, for the initial condition $N(\tau_0) = 0$. It is assumed that the beginning of inflation was at some time $\tau_i > 0$, rather than at $\tau_0 = 0$. The time of the beginning of inflation $\tau_i$ is obtained from the requirement $N(\tau_f) - N(\tau_i) = N$.

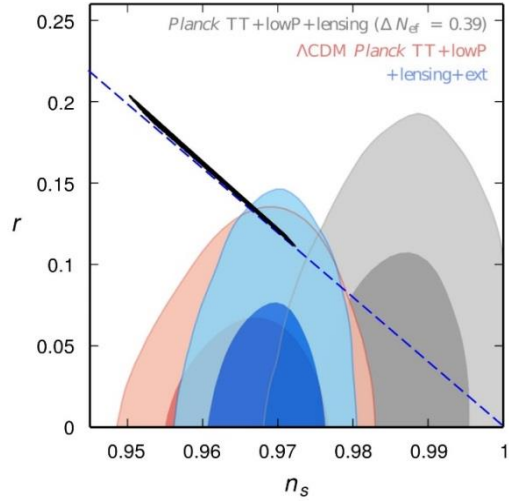

**Fig. 1** Observational parameters of inflation ($n_s$,$r$) with observational constraints from Planck Collaborations 2015 (Ade et al., 2016a). Dashed (blue) line shows the preliminary analytical results in slow-roll approximation, for $\varepsilon_2(\tau_i) \approx 2\varepsilon_1(\tau_i)$. Each point corresponds to one pair of ($N$,$\kappa$).

We obtain the distribution of numerical values for the observational parameters $n_s$ and $r$ from Eqs. (37) and (38). The simulation is performed for randomly chosen combinations of $N$ and $\kappa$, in the range $45 < N < 75$ and $1 < \kappa < 10$. The numerical results are superimposed on the observational data taken from the Planck Collaboration (Ade et al., 2016a; Ade et al., 2016b) and presented in Fig. 1.

## 7. CONCLUSION

We studied the inflationary scenario in the context of the RSII model extended to include matter in the bulk. The dynamics of the model is determined by the motion of a D3-brane in the $AdS_5$ bulk. The resulting effective model is described by tachyon field theory. Out of the wide class of tachyon potentials, we studied in more detail the model based on the exponential potential. It is a reasonably simple and very instructive model.

Observational parameters of inflation were calculated numerically, for a range of numbers of e-folds $N$ and the dimensionless coupling constant $\kappa$. In spite of a slight agreement with the Planck data (Ade et al., 2016a; Ade et al., 2016b), available at the moment when the numerical calculation was done, this agreement is not quite satisfactory. A comparison with our previous results for the same potential in a different framework (Milošević, 2016; Milošević, 2019) shows that the RSII model with matter in the bulk gives a significantly better agreement. One should bear in mind that our model is based on brane dynamics described by a simple exponential potential with only one free parameter. Anyhow, the brane world cosmology significantly improved the results obtained for this potential in the standard FRWL cosmology mentioned above.

We would like to mention here some approximative results. If one keeps higher derivatives of $\chi$ with the respect to the field $\theta$ in the slow-roll regime, i.e. $\chi_{,\theta\theta}$ and $\chi_{,\theta\theta\theta}$ in Eq. (26), then Eqs. (25)-(26) lead to $\varepsilon_2(\tau_i) \approx 2\varepsilon_1(\tau_i)$. Further, applying this approximation to Eqs. (37)-(38), it leads to linear dependence between the observational $n_s$ and $r$, which corresponds to the highest density of the numerical results (dashed blue line in Fig. 1). The corresponding approximative relation for $\varepsilon_1$ and $\varepsilon_2$ and its implementation in our model for more suitable tachyon potentials (as cosh and similar) in the context of extended RSII cosmology are subject of an ongoing study. The results will be published and discussed in detail elsewhere.

**INFLATORNI RSII MODEL SA MATERIJOM U BULK-U
I TAHIONSKO POLJE
SA EKSPONENCIJALNIM POTENCIJALOM**

*U ovom radu razmatramo tahionski kosmološki model zasnovan na dinamici 3-brane u drugom Randall-Sundrum (RSII) modelu, koji je proširen uvođenjem materije u bulk-u. Prisustvo materije u bulk-u menja warp faktor, što dovodi do modifikacije inflatorne dinamike. Dodatna brana se efektivno ponaša kao tahion. Za slučaj eksponencijalnog potencijala tahionskog polja numerički su izračunati posmatrački parametri inflacije: skalarni spektralni indeks ($n_s$) i tenzor-skalar odnos ($r$).*

Ključne reči: *DBI lagranžijan, tahionska inflacija, uopšteni RSII model*